\documentclass[11pt]{amsart}
\usepackage{amsmath}
\usepackage{amsfonts}
\usepackage{amssymb}
\newtheorem{thm}{Theorem}

\newtheorem{cor}[thm]{Corollary}

\newtheorem{lemma}[thm]{Lemma}
\newtheorem{prop}[thm]{Proposition}

\newtheorem{exam}[thm]{Example}
\newtheorem{defn}[thm]{Definition}

\newcommand{\Tr}{{\rm Tr}}
\newcommand{\bb}[1]{\mathbb{#1}}
\newcommand{\cl}[1]{\mathcal{#1}}

\begin{document}

\title[Computing Stabilized Norms for Quantum Operations]{Computing Stabilized Norms for Quantum Operations via the Theory of Completely Bounded Maps}

\author[N. Johnston, D.~W.~Kribs, V.~I.~Paulsen]{Nathaniel Johnston$^1$, David~W.~Kribs$^{1,2}$, and Vern~I.~Paulsen$^3$}
\address{$^1$Department of Mathematics \& Statistics, University of Guelph,
Guelph, ON, Canada N1G 2W1}
\address{$^2$Institute for Quantum Computing, University of Waterloo, Waterloo, ON, Canada
N2L 3G1}
\address{$^3$Department of Mathematics, University of Houston,
Houston, Texas 77204-3476, U.S.A.}
\email{njohns01@uoguelph.ca}\email{dkribs@uoguelph.ca}
\email{vern@math.uh.edu}


\begin{abstract}
The diamond and completely bounded norms for linear maps play an
increasingly important role in quantum information science,
providing fundamental stabilized distance measures for differences
of quantum operations. Based on the theory of completely bounded
maps, we formulate an algorithm to compute the norm of an
arbitrary linear map. We present an implementation of the
algorithm {\it via} Maple, discuss its efficiency, and consider
the case of differences of unitary maps.
\end{abstract}

\maketitle


\section{Introduction}

The need for physically significant and computable distance
measures for quantum operations and channels is of fundamental
importance in quantum information science \cite{NC00}. Most
importantly, it is often necessary to determine how far apart two
quantum operations, represented by completely positive maps, are
from each other in some meaningful sense. The {\it diamond norm}
was introduced in \cite{Kit97} for this purpose. It arises from
physical considerations and satisfies the important stability
property desired for such measures \cite{AKN97,GLN05}.
Interestingly, the diamond norm is intimately related to the {\it
norm of complete boundedness}, a notion that has been studied in
operator theory for different reasons over the past four decades
\cite{Pa}. On finite dimensional Hilbert space, every linear map
has a finite completely bounded (CB) norm. Thus, CB maps are
precisely the linear maps in the finite dimensional case. In
operator theory, CB maps are the natural maps between certain
objects called operator spaces. Computing the norms of CB maps
between certain operator spaces introduced in \cite{Pi98} has
provided the impetus for recent progress on multiplicativity
conjectures for quantum channels \cite{DJKR05,DHLW07,Hay07,Win07}.
CB maps and norms have also arisen in a wide variety of other
recent investigations in quantum information science, including
\cite{Jen07,JSS04,KKS07,KSW07,PWPVJ07,SL07,Wat05}, though the CB
terminology has not always been used.

In this paper, based on the classical and contemporary theory of
completely bounded maps, we formulate an algorithm to compute
completely bounded and diamond norms for arbitrary linear maps on
finite dimensional Hilbert space. Along the way, we also provide a
brief introduction to completely bounded map theory, including the
generalized Stinespring theorem and Choi-Kraus representation for
such maps. We then present an implementation of the algorithm {\it
via} Maple. We also discuss the algorithm's efficiency, and note
how it is potentially optimal. Our approach to computing these
norms is distinct from other known approaches, such as the use of
semidefinite programming \cite{Watprivate}.

In the next section we recall basic properties of the diamond and
completely bounded norm, showing how to interpolate between the
two. This is followed by the introduction to completely bounded
maps; our presentation here is motivated by that of \cite{Pa}. In
the penultimate section we describe the theoretical formulation of
the algorithm, and apply it to derive a geometric formula for the
case of differences of unitary maps. In the final section we
exhibit code for the Maple implementation of the algorithm, giving
an explanation of each subroutine.

\section{Linear Maps and Stabilized Norms}

We shall write $M_n$ for the set of $n\times n$ complex matrices,
and regard it as the set of operators acting on an $n$-dimensional
Hilbert space represented as matrices in a given orthonormal
basis. {\it Quantum operations} or {\it channels} are represented
by linear maps $\phi : M_n \rightarrow M_k$ that are completely
positive and trace preserving (in the Schrodinger picture) or
unital (in the Heisenberg picture). The dual map $\phi^\dagger :
M_k \rightarrow M_n$ is defined via the Hilbert-Schmidt inner
product $\Tr(\phi(A)B) = \Tr(A\, \phi^\dagger(B))$.

In quantum information one is often interested in properties of
differences $\phi - \psi$ between pairs of quantum operations.
Such a difference is still a linear map (a ``superoperator''),
though not necessarily completely positive. In fact, every linear
map can be decomposed as a linear combination of at most four
completely positive maps. This parallels the corresponding
statement about general operators and positive operators on
Hilbert space, though the proof is more delicate \cite{Witt,Pa}.

The 1-norm of a linear map $\phi$ is given by $||\phi||_1 =
\sup_{||X||_1\leq 1} ||\phi(X)||_1,$ where $||X||_1 = \Tr |X|$.
The operator norm of $\phi$ is $||\phi|| = \sup_{||X||\leq 1}
||\phi(X)||,$ where $||X||= \sup_{||\xi||\leq 1} ||X\xi||$, and
$\xi$ ranges over the unit ball of the domain Hilbert space for
$X$. Every norm $|||\phi|||$ defines a distance measure
$d(\phi,\psi) = |||\phi - \psi|||$. Neither of the distance
measures defined by these norms satisfies the {\it stabilization
property} for distance measures of superoperators
\cite{Kit97,AKN97,GLN05}:
$$d(id_m\otimes\phi,id_m\otimes\psi)= d(\phi,\psi) \quad \forall
m\geq 1. $$ This property implies that the distance between
quantum operations is unaffected by any ancillary quantum system
that is independent of the original system.

The diamond norm is defined in \cite{Kit97,AKN97} through partial
traces, but it is shown to be equivalent to the following
quantity.

\begin{defn}
For a linear map $\phi: M_n \rightarrow M_k$, define
\[
||\phi||_\diamond \,\, = \,\, || id_n \otimes \phi ||_1.
\]
\end{defn}

Though not obvious, the stabilization property is satisfied by
$||\cdot||_\diamond$. One way to see this is through a connection
with the completely bounded norm, to which we now turn.

\begin{defn}
For a linear map $\phi: M_n \rightarrow M_k$, define
\[
||\phi||_{cb} \,\, = \,\, \sup_{m\geq 1} || id_m \otimes \phi ||,
\] which we call the {\bf completely bounded norm} or the {\bf cb-norm.}
\end{defn}

As a convenience, we adopt the notation $\phi_m \equiv id_m
\otimes \phi$. It is easily checked that $\|\phi_m\| \le
\|\phi_{m+1}\|$ and $\|\phi_m \| \le m \|\phi\|$ \cite[Chapter
1]{Pa}. Note also that $||{\mathcal U}\, \phi ||_{cb} = ||\phi
||_{cb} = ||\phi \, {\mathcal U}||_{cb}$ for every unitarily
implemented map $\mathcal U (\cdot) = U (\cdot) U^\dagger$.
Furthermore, we have $||\phi_1 \otimes \phi_2||_{cb} =
||\phi_1||_{cb}||\phi_2||_{cb}$, and $||{\rm id}||_{cb}=1$. The
identification in Theorem~\ref{interpolate} shows the
corresponding properties hold for the diamond norm.

It is possible to relate the completely bounded norm and the
diamond norm as follows. We first note that by a theorem of Smith
\cite{Smi83,Pa} the cb-norm stabilizes in the sense that for a map
$\phi: M_n \to M_k$ we have that $||\phi||_{cb} = ||id_k \otimes
\phi||.$  Then we make use of the duality relationship
\cite{BrRo87} given by $||\phi || = \sup_{||X||_1\leq 1}
||\phi^\dagger(X)||_1$, to obtain
\begin{eqnarray*}
||\phi||_{cb} &=& ||id_k \otimes \phi||  = \sup_{||X||_1\leq 1}
||(id_k\otimes\phi^\dagger)(X)||_1 = ||\phi^\dagger||_\diamond,
\end{eqnarray*}
since $\phi^\dagger: M_k \to M_n.$

Thus, we have the following, which also includes an upper bound
\cite{Smi83}.

\begin{thm}\label{interpolate} Let $\phi: M_n \to M_k,$ be a linear map, then $$\|\phi\|_{cb} = \|\phi^\dagger\|_\diamond = \|\phi_k\| \le k
\|\phi\|.$$
\end{thm}

In summary, using the fact that these maps appear as dual pairs,
we see that for $\psi: M_m \to M_j, \|\psi\|_\diamond =
\|\psi^\dagger\|_{cb} \le m \|\psi\|$ and that the stability of
the diamond norm \cite{Kit97,AKN97} is the dual version of Smith's
stability for the cb-norm \cite{Smi83}. A more refined general
upper bound is discussed in the next section.

\section{Completely Bounded Map Primer}

We next give a compressed introduction to completely bounded maps
on arbitrary operator spaces. As noted previously, in the finite
dimensional case CB maps on $M_n$ are precisely the linear maps
$\phi: M_n \rightarrow M_k$. However, the important structural
results reviewed in this section are best viewed in the more
general setting of CB maps as in \cite{Pa}. In any event, if the
reader wishes to move directly to the algorithm, this section can
be skipped save for the structural result Theorem~\ref{cbrepn} and
the norm description of Corollary~\ref{cbrepncor}.

Given a (separable) Hilbert space $\cl H$, we denote the set of
(bounded) linear operators on $\cl H$ by $B(\cl H)$. Given
operators, $T_{i,j} \in B(\cl H), 1 \le i \le m, 1 \le j \le n,$
we identify the $m \times n$ matrix of operators, $(T_{i,j})$,
with an operator from $\cl H^{(n)}= \cl H \oplus \ldots \oplus \cl
H$ (n copies) to $\cl H^{(m)} = \cl H \oplus \ldots \cl H$ (m
copies) by regarding vectors in these spaces as columns and
performing matrix multiplication. That is, we identify
$M_{m,n}(B(\cl H)) \equiv B(\cl H^{(n)}, \cl H^{(m)})$. This
endows $M_{m,n}(B(\cl H))$ with a norm and this collection of
norms on $B(\cl H)$ is often referred to as the set of {\it matrix
norms} on $B(\cl H).$

\begin{defn} Let $\cl H$ be a Hilbert space and let $\cl M \subseteq B(\cl H)$ be a subspace. Then the inclusion, $M_{m,n}(\cl M) \subseteq M_{m,n}(B(\cl H))$ endows this vector space with a collection of matrix norms and we call, $\cl M$, together with this collection of matrix norms on $M_{m,n}(\cl M)$ a (concrete) {\bf operator space.} When $m=n,$ we set $M_n(\cl M)= M_{n,n}(\cl M).$
\end{defn}

Thus, an operator space carries not just an inherited norm
structure, but these additional matrix norms.

For basic properties of C*-algebras we point the reader to
\cite{Dav96a}. C*-algebras are defined abstractly, but every
abstract algebra is isomorphic to a concrete C*-algebra given by a
subalgebra of some $B(\cl H)$ that is closed under both the
operator norm ($||\cdot||$) and adjoint ($\dagger$) operation. If
$\cl A$ is any C*-algebra and $\pi: \cl A \to B(\cl H)$ is a
one-to-one $\dagger$-homomorphism (and hence an isometry), then
the collection of norms on $M_{m,n}(\pi(\cl A))$ is independent of
the particular representation $\pi,$ and hence, the operator space
structure of a C*-algebra is independent of the particular
(faithful) representation. Hence, each subspace $\cl M \subseteq
\cl A$ is also endowed with a particular collection of matrix
norms and so we also refer to a subspace of a C*-algebra as an
operator space, when we wish to emphasize its matrix norm
structure. We now give the definition of a completely bounded map
in the general operator space setting.

\begin{defn}
Given a C*-algebra $\cl A$, an operator space $\cl M \subseteq \cl
A,$ and a linear map, $\phi: \cl M \to B(\cl H),$ we define
$\phi_n: M_n(\cl M) \to M_n(B(\cl H))$ by $\phi_n((a_{i,j})) = (
\phi(a_{i,j})).$ We call $\phi$ {\bf completely bounded}, if
$$\|\phi \|_{cb} \equiv \sup_n \| \phi_n \|,$$ is finite. Here
$||\phi_n || = \sup \big\{ ||\phi_n(A)|| \, : \, A\in M_n(\cl M)
,\, ||A||\leq 1 \big\}$.
\end{defn}

More generally, any time that $\cl M$ and $\cl N$ are two spaces,
both endowed with a family of matrix norms, then one can define
the completely bounded norm of a map $\phi : \cl M \rightarrow \cl
N$ in analogy with the above definition.

Recalling the upper bound of Theorem~\ref{interpolate}, for maps
whose domain is $M_n$ and range an arbitrary operator space, a
result of Haagerup shows that, in general, $\|\psi\|_{cb} \ne
\|\psi_m \|$, no matter how large one takes $m$ \cite[p. 114]{Pa},
but we do have an upper bound. This result is not explicitly in
the literature so we provide a proof below, that uses some
concepts we will introduce in Section 4 and, perhaps, illustrates
their utility. For now, it is enough to know that given a finite
dimensional normed space $X$, there exists a constant $\alpha(X)$
called the {\em alpha constant} of the space, with the property
that $$\|\phi\|_{cb} \le \alpha(X) \|\phi\|$$ for any map with
domain an operator space that is isometrically isomorphic to $X$
as normed spaces. Given two finite dimensional normed spaces $X,Y$
of the same dimension one has
$$\alpha(X) \le d(X,Y) \alpha(Y),$$ where $d(X,Y)$ denotes the
{\em Banach-Mazur distance} between the spaces. These concepts and
results can be found in \cite{Pa2}.

\begin{thm} Let $\cl M$ be an operator space and let $\phi: M_n \to \cl M$ be a linear map, then $\|\phi\|_{cb} \le n \sqrt{n} \|\phi\|.$
\end{thm}
\begin{proof} Let $\|X\|_2$ denote the Hilbert-Schmidt norm of a matrix $X$. Since $\|X\| \le \|X\|_2 \le \sqrt{n} \|X\|,$
the Banach-Mazur distance satisfies $d(M_n, \bb C^{n^2}) \le
\sqrt{n}.$

Hence, $\alpha(M_n) \le d(M_n, \bb C^{n^2}) \alpha(\bb C^{n^2})
\le \sqrt{n} \alpha(\bb C^{n^2}).$ Finally, it is shown in
\cite{Pa2}, that for Euclidean space, $\alpha(\bb C^m) \le
\sqrt{m}$, from which the result follows.
\end{proof}

Combining this result with Theorem~\ref{interpolate} we have:

\begin{cor} Let $\phi: M_n \to M_k$ be a linear map, then $$\|\phi\|_{cb} \le \min\{k,
\sqrt{n^3}\}\, ||\phi||.$$
\end{cor}

We next recall the abstract definition of completely positive
maps.

\begin{defn} If $\cl A$ is a unital C*-algebra, then a $\dagger$-closed subspace $\cl S \subseteq \cl A$ such that $1 \in \cl S$, is called an {\bf operator system.}
\end{defn}

Thus, operator systems are operator spaces and have matrix norms.
But the additional hypotheses guarantee that if we let $\cl A^+$
denote the positive elements of the C*-algebra, then $\cl S$ is
the span of $\cl S^+ \equiv \cl S \cap \cl A^+,$ which is a cone
in $\cl S.$ We also have that $M_n(\cl S)$ is the span of $M_n(\cl
S)^+= M_n(\cl S) \cap M_n(\cl A)^+.$ The vector spaces, $M_n(\cl
S)$ together with the cones $M_n(\cl S)^+$ is often referred to as
the {\it matrix ordering} on $\cl S.$

\begin{defn} Given a unital C*-algebra $\cl A, \cl S \subseteq \cl A$ and a map $\phi: \cl S \to B(\cl H),$ we call $\phi$ {\bf completely positive,} provided that $\phi_n$ is positive for all $n,$ that is provided that $(a_{i,j}) \in M_n(\cl S)^+$ implies that $(\phi(a_{i,j})) \in M_n(B(\cl H))^+.$
\end{defn}

\begin{defn} Given a C*-algebra $\cl A$ and $\cl M \subseteq \cl A$ an operator space, we set $\cl M^*= \{ a^\dagger: a \in \cl M \},$ which is another operator space.
If $\phi: \cl M \to B(\cl H),$ is a linear map, then we define
$\phi^*: \cl M^* \to B(\cl H)$ by $\phi^*(b) =
\phi(b^\dagger)^\dagger,$ which is another linear map.
\end{defn}

The following objects allow one to relate much of the theory of
completely bounded maps to the more familiar theory of completely
positive maps.

\begin{defn} Let $\cl A$ be a unital C*-algebra and let $\cl M \subseteq \cl A$ be an operator space, then we define an operator system $\cl S_{\cl M} \subseteq M_2(\cl A),$ by
$$\cl S_{\cl M} \equiv \Big\{ \begin{pmatrix} \lambda 1 & a\\ b^\dagger & \mu 1 \end{pmatrix} : \lambda \in \bb C, \mu \in \bb C, a \in \cl M, b \in \cl M \Big\}.$$
\end{defn}

Details for the next six results can be found in \cite{Pa}. We
briefly sketch some of the ideas of the proofs to help indicate
the interplay between completely bounded maps and completely
positive maps.

\begin{thm} Let $\cl A$ be a unital C*-algebra, $\cl M \subseteq \cl A$ be an operator space and let $\phi: \cl M \to B(\cl H)$ be linear. Then $\|\phi\|_{cb} \le 1$ if and only if $\Phi: \cl S_{\cl M}: \to M_2(B(\cl H))$ is completely positive, where $$\Phi\big(\begin{pmatrix} \lambda 1 & a\\ b^\dagger & \mu 1 \end{pmatrix}\big) = \begin{pmatrix} \lambda I_{\cl H} & \phi(a) \\ \phi(b)^\dagger & \mu I_{\cl H} \end{pmatrix}.$$
\end{thm}

In particular, using this identification of completely contractive
($||\phi||_{cb} \leq 1$) maps with ``corners'' of unital
completely positive maps, one extension theorem:

\begin{thm}[Arveson's Extension Theorem] Let $\cl S \subseteq \cl A$ be an operator system and let $\phi: \cl S \to B(\cl H)$ be completely positive. Then there exists a completely positive map $\psi: \cl A \to B(\cl H)$ that extends $\phi$, that is, such that $\psi(a) = \phi(a)$ for every $a \in \cl S.$
\end{thm}

quickly yields another:

\begin{thm}[Wittstock's Extension Theorem]  Let $\cl M \subseteq \cl A$ be an operator space and let $\phi: \cl M \to B(\cl H)$ be completely bounded. Then there exists a completely bounded map $\psi: \cl A \to B(\cl H)$ that extends $\phi$ and satisfies $\|\psi\|_{cb} = \|\phi\|_{cb}.$
\end{thm}

To obtain the second from the first, one first scales $\phi$ so
that $\|\phi \|_{cb}=1,$ then applies Arveson's Theorem to extend
$\Phi: \cl S_{\cl M} \to M_2(B(\cl H)),$ to $\Psi: M_2(\cl A) \to
M_2(B(\cl H)),$ and then lets $\psi$ be the corresponding (1,2)
matrix corner of $\Psi.$

A fundamental result for quantum information is Stinespring's
classical representation theorem for completely positive maps.

\begin{thm}[Stinespring's Representation Theorem]  Let $\cl A$ be a unital C*-algebra and let $\phi: \cl A \to B(\cl H)$ be a completely positive map, then there exists a Hilbert space $\cl K,$ a bounded operator $V: \cl H \to \cl K$ and a unital $\dagger$-homomorphism, $\pi: \cl A \to B(\cl K)$ such that $\phi(a) = V^\dagger\pi(a)V,$ for every $a \in \cl A.$
\end{thm}

Note that in Stinespring's theorem, we also have that
$\|\phi\|_{cb}= \|\phi(1)\|= \|V^\dagger V\|= \|V\|^2.$

This form of the Stinespring Theorem is less common in quantum
information, but the more standard forms can be readily obtained.
Suppose $\phi : M_n \rightarrow M_k$ is a completely positive
unital map. As every representation of $M_n$ is unitarily
equivalent to a multiple of the identity representation
\cite{Dav96a}, $\pi$ can be assumed to be of the form $\pi (a) =
I_E \otimes a$, where $I_E$ is the identity operator on a suitable
dilation Hilbert space $E$. Further, as $\phi$ is unital we have
$I = \phi(I) = V^\dagger V$, and hence $V$ is an isometry. Thus we
may write $\phi$ as $\phi(a) = V^\dagger (I_E \otimes a) V$. The
dual of this equation yields the familiar partial trace form for a
quantum channel, $\phi^\dagger(a) = \Tr_E (VaV^\dagger)$.

In a similar fashion to the extension theorem, Stinespring's
Theorem can be extended to completely bounded maps.

\begin{thm}[The Generalized Stinespring Theorem]  Let $\cl A$ be a unital C*-algebra and let $\phi: \cl A \to B(\cl H)$ be a completely bounded map, then there exists a Hilbert space $\cl K,$ bounded operators $V: \cl H \to \cl K, W: \cl H \to \cl K$ and a unital $\dagger$-homomorphism, $\pi: \cl A \to B(\cl K),$ such that $\|\phi\|_{cb}= \|V\|\|W\|$ and $\phi(a) = V^\dagger\pi(a)W,$ for every $a \in \cl A.$
\end{thm}

In the finite dimensional case of a completely bounded map $\phi:
M_n \rightarrow M_k$, the corresponding canonical forms look like
$\phi(a) = V^\dagger (I_E \otimes a) W$ and $\phi^\dagger(a) =
\Tr_E(W a V^\dagger)$.

The generalization of Stinespring's theorem to completely bounded
maps also yields the following ``polar form'' for completely
bounded maps. To motivate this result note that for operators, if
we set $|T|= \sqrt{T^\dagger T},$ then $\begin{pmatrix} |T^\dagger| & T\\
T^\dagger & |T|
\end{pmatrix}$ is positive.

\begin{cor} Let $\cl A$ be a unital C*-algebra and let $\phi: \cl A
  \to B(\cl H)$ be completely bounded, then there exists completely
  positive maps, $\phi_1, \phi_2: \cl A \to B(\cl H),$ with $\| \phi_1(1)\|= \|\phi_2(1)\|= \|\phi\|_{cb},$ such that $\Phi:M_2(\cl A) \to M_2(B(\cl H))$ is completely positive, where $$\Phi( \begin{pmatrix} a & b\\c & d \end{pmatrix}) = \begin{pmatrix} \phi_1(a) & \phi(b) \\ \phi^*(c) & \phi_2(d) \end{pmatrix}.$$
\end{cor}

When $\|\phi\|_{cb} \le 1,$ then $\phi_1,\phi_2$ can both be taken
to be unital completely positive maps. Thus, the above corollary
is another example of the meta-theorem that completely contractive
maps are the ``corners'' of unital completely positive maps.

The generalized Stinespring theorem, unfortunately, has no good
uniqueness criteria, unlike the usual Stinespring theorem. The
difficulty stems from the fact that the two completely positive
maps, $\phi_1, \phi_2$ are not uniquely determined by $\phi.$
Generally, there are many possible extensions of the completely
positive map $\Phi$ from the operator system $\cl S_{\cl A}$ to
$M_2(\cl A)$ and this allows for a great deal of non-uniqueness.
Ostensibly this follows from the fact that the Hahn-Banach Theorem
plays a key role in the proof. Thus, in particular, the above
``polar form'' of a completely bounded map is not unique.

Just as one obtains the Choi-Kraus representation of completely
positive maps from $M_n$ to $M_k$ by specializing Stinespring's
theorem to these algebras, one obtains a similar representation of
completely bounded maps.

\begin{thm}[Choi-Kraus Representation Theorem \cite{Cho75,Kra71}] Let $\phi: M_n \to M_k$ be completely positive, then there exists matrices, $A_i \in M_{n,k}, 1 \le i \le nk,$ such that $\phi(X) = \sum_{i} A_i^\dagger XA_i.$
\end{thm}

\begin{thm}[CB Representation Theorem]\label{cbrepn}
Let $\phi: M_n \to M_k$ be a linear map. Then there exists
matrices, $A_i \in M_{k,n}, 1 \le i \le m,$ and matrices $B_i \in
M_{n,k}, 1 \le i \le m,$ such that
\begin{eqnarray}\label{c-kform}
\phi(X) = \sum_i A_iXB_i,
\end{eqnarray}
with $\|\phi\|_{cb}^2 = \|\phi^\dagger \|_\diamond^2  = \| \sum_i
A_iA_i^\dagger \| \| \sum_i B_i^\dagger B_i \|$ and $m \le nk.$
\end{thm}

It is important to understand the difference between $\phi^*$ and
the usual dual map $\phi^\dagger$ considered in quantum
information, so let us dwell on this point for a moment with $\cl
M= M_n, B(\cl H)=M_k,$ and $\phi:M_n \to M_k$.   In terms of
Choi-Kraus operation elements, if $A \in M_{k,n}, B \in M_{n,k},$
and $\phi:M_n \to M_k,$ is defined by $\phi(X)= AXB,$ then
$\phi^*: M_n \to M_k$ is given by $\phi^*(X)= B^*XA^*,$ while
$\phi^{\dag}: M_k \to M_n$ is given by $\phi^{\dag}(Y)= BYA$, and
the obvious generalization holds true if $\phi$ is given by a sum
of such terms.

Another difference between $\phi^*$ and $\phi^\dagger$ arises when
considering the CB norm. It is easily checked that $\|\phi_n \| =
\| (\phi^*)_n\|$ and hence that $\|\phi\|_{cb}= \|\phi^*\|_{cb}.$
On the other hand, $||\phi||_{cb}$ and $||\phi^\dagger||_{cb}$ can
be different. For instance, in the case of a completely positive,
trace preserving map $\phi$, the dual $\phi^\dagger$ is unital
($\phi^\dagger(I)=I$), so that
$||\phi^\dagger||_{cb}=||\phi^\dagger(I)|| = 1$, whereas
$||\phi||_{cb}=||\phi(I)||$ could be larger or smaller.

We shall call {\em any} representation $\phi(X) = \sum_i A_iXB_i$
a {\it generalized Choi-Kraus representation.} Note that if we
have any generalized Choi-Kraus representation of $\phi$ then,
$$\phi(X) = (A_1, \ldots , A_m) \begin{pmatrix} X & 0 & \dots & 0 \\ 0
  & X & \dots & 0 \\ \vdots & \vdots & \ddots & \vdots \\ 0 & 0 &
  \dots & X \end{pmatrix}\begin{pmatrix} B_1\\ \vdots \\ B_m \end{pmatrix},$$ where the term in the middle represents the $m \times m$ block diagonal matrix each of whose blocks is $X$
and hence,
$$\|\phi\|_{cb} \le \|(A_1,..., A_m)\| \|\begin{pmatrix} B_1\\ \vdots \\ B_m \end{pmatrix} \| = \| \sum_i A_iA_i^\dagger \|^{1/2} \| \sum_i B_i^\dagger B_i \|^{1/2},$$
which explains the asymmetry in the roles of the A's and B's.

This also leads to the following very useful result.

\begin{cor}\label{cbrepncor}
Let $\phi:M_n \to M_k$ be a linear map, then $$\|\phi\|_{cb} =
\|\phi^\dagger \|_\diamond = \inf \Big\{ \| \sum_i A_iA_i^\dagger
\|^{1/2} \| \sum_i B_i^\dagger B_i \|^{1/2} \Big\},$$ where the
infimum is taken over all generalized Choi-Kraus representations
of $\phi.$
\end{cor}


\section{Computation and Estimation  of the CB/$\Diamond$ Norm}

In this section we present the theoretical formulation of the
algorithm and use it to derive a geometric formula for mixtures of
pairs of unitary maps. In the case of a completely positive map
$\phi$, Theorem~\ref{cbrepn} coupled with the Choi-Kraus
representation theorem shows that the CB norm of $\phi$ is exactly
the operator norm $||\phi(I)||$. For completeness we provide the
direct, elementary proof of this fact from \cite{Pa} with no
restriction on the domain of the map.

\begin{lemma} \label{lem:posmatrix}
Let $P$ and $A$ be operators on some Hilbert space $\mathcal{H}$
with $P$ positive. Then $\left( \begin{matrix}P & A \\ A^{*} & P
\end{matrix}\right) \geq 0$ implies that $\left\| A \right\| \leq
\left\| P \right\|$. Furthermore, if $P$ is the identity operator
then the converse also holds.
\end{lemma}

\begin{proof}
    To show the forward implication, note that if $\left( \begin{matrix}P & A \\ A^{*} & P \end{matrix}\right) \geq 0$ then it follows that $\left\langle \left( \begin{matrix}P & A \\ A^{*} & P \end{matrix}\right) \left(\begin{matrix} x \\ -y \end{matrix}\right) \bigg{|} \left(\begin{matrix} x \\ -y \end{matrix}\right) \right\rangle \geq 0$ $\forall x,y \in \mathcal{H} \text{ s.t. } \left\| x \right\| = \left\| y \right\| = 1$. Thus, $\left\langle Px | x \right\rangle + \left\langle Py | y \right\rangle \geq \left\langle Ay | x \right\rangle + \left\langle x | Ay \right\rangle = 2 Re\left(\left\langle Ay | x \right\rangle\right)$.  Also, the Cauchy-Schwarz Inequality tells us that $\left\langle Px | x \right\rangle + \left\langle Py | y \right\rangle \leq \left\| Px \right\| + \left\| Py \right\| \leq 2\left\| P \right\|$ since $\left\| x \right\| = \left\| y \right\| = 1$. Thus, $\left\| P \right\| \geq Re\left(\left\langle Ay | x \right\rangle\right) \forall x,y \in \mathcal{H} \text{ s.t. } \left\| x \right\| = \left\| y \right\| = 1$. A little thought reveals that this is equivalent to $\left\| P \right\| \geq \left\langle Ay | x \right\rangle \forall x,y \in \mathcal{H} \text{ s.t. } \left\| x \right\| = \left\| y \right\| = 1$, which immediately implies that $\left\| A \right\| \leq \left\| P \right\|$.

    To show the converse when $P = I$ is the identity operator, we prove by contradiction by assuming that $\left\| A \right\| \leq \left\| I \right\| = 1$ and that $\exists x,y \in \mathcal{H}$ such that $\left\langle \left( \begin{matrix}I & A \\ A^{*} & I \end{matrix}\right) \left(\begin{matrix} x \\ -y \end{matrix}\right) \bigg{|} \left(\begin{matrix} x \\ -y \end{matrix}\right) \right\rangle < 0$. It then follows that $\left\| x \right\|^2 + \left\| y \right\|^2 < \left\langle Ay | x \right\rangle + \left\langle x | Ay \right\rangle$.

    The Cauchy-Schwarz Inequality tells us that $\left\langle Ay | x \right\rangle + \left\langle x | Ay \right\rangle \leq \left\| Ay \right\| \left\| x \right\| + \left\| x \right\| \left\| Ay \right\| \leq 2 \left\| A \right\| \left\| x \right\| \left\| y \right\| \leq 2 \left\| x \right\| \left\| y \right\|$. Thus, $\left\| x \right\|^2 + \left\| y \right\|^2 < 2 \left\| x \right\| \left\| y \right\|$. This, however, is impossible since $\forall x, y \in \mathcal{H}$ it is true that $(\left\| x \right\|^2 + \left\| y \right\|^2)^2 = (\left\| x \right\|^2 - \left\| y \right\|^2)^2 + 4 \left\| x \right\|^2 \left\| y \right\|^2 \geq 4 \left\| x \right\|^2 \left\| y \right\|^2$, so $\left\| x \right\|^2 + \left\| y \right\|^2 \geq 2 \left\| x \right\| \left\| y \right\|$, completing the contradiction.
\end{proof}

\begin{thm} \label{thm:main}
Let $\mathcal{S} \subseteq \mathcal{A}$ be an operator system, let
$\mathcal{B}$ be a C*-algebra, and let $\phi : \mathcal{S}
\rightarrow \mathcal{B}$ be a completely positive map. Then $\phi$
is completely bounded and $\left\| \phi \right\|_{cb} = \left\|
\phi \right\| = \left\| \phi (1) \right\|$.
\end{thm}

\begin{proof}
    First note that $\left\| \phi (1) \right\| \leq \left\| \phi \right\| \leq \left\| \phi \right\|_{cb}$, so we need only show that $\left\| \phi \right\|_{cb} \leq \left\| \phi (1) \right\|$.

    Fix $n$ and let $A \in M_{n}(\mathcal{S})$ be such that $\left\| A \right\| \leq 1$, and let $I_n$ be the unit of $M_{n}(\mathcal{A})$. Then Lemma~\ref{lem:posmatrix} tells us that $\left( \begin{matrix}I_n & A \\ A^{*} & I_n \end{matrix}\right) \geq 0$. Since $\phi$ is completely positive, it then follows that $\phi_{2n} \left( \begin{matrix} I_n & A \\ A & I_n \end{matrix}\right) = \left( \begin{matrix}\phi_n \left( I_n \right) & \phi_n \left( A \right) \\ \phi_n \left( A \right)^{*} & \phi_n \left( I_n \right) \end{matrix}\right) \geq 0$. Making use of Lemma~\ref{lem:posmatrix} again shows us that $\left\| \phi_n \left( A \right) \right\| \leq \left\| \phi_n \left( I_n \right) \right\| = \left\| \phi \left( 1 \right) \right\|$. Since this inequality holds for any such $A$, the proof is complete.
\end{proof}

\subsection{The Algorithm.}

We now turn to the problem of actually computing the norm of a
linear map $\phi:M_n \to M_k.$ By the above results we know that
to compute the cb-norm we need to do a minimization over all
generalized Choi-Kraus representations. This turns out to be
somewhat more attainable than might be imagined and we present an
algorithm for computing the cb/$\diamond$-norm of such maps. We
first describe the algorithm and then justify it later.

We assume that we are given a map $\phi: M_n \to M_k,$ some
generalized Choi-Kraus representation $\phi(X) = \sum_{i=1}^m
A_iXB_i$ and we wish to compute $\|\phi\|_{cb} =
||\phi^\dagger||_\diamond.$

\bigskip

{\bf Step 1.}  Find a basis, $\{ C_1,...,C_l \}$ for the span of
$\{B_1,..., B_m \}$ and express $B_i = \sum d_{i,j}C_j$

\bigskip

{\bf Step 2.} Using the expressions for each $B_i$ as a linear
combination of $C_j$ we may re-write $\phi(X) = \sum_{j=1}^l
D_jXC_j.$ In fact, we have $\phi(X) = \sum_i A_iX(\sum_j
d_{i,j}C_j) = \sum_j ( \sum_i d_{i,j}A_i)XC_j.$  Thus,
$$D_j = \sum_i d_{i,j} A_i.$$

\bigskip

{\bf Step 3.} Find a  basis $\{E_1,..., E_p \}$ for the span of
$\{ D_1,...,D_l \}$, express each $D_j$ as a linear combination,
and repeat Step 2 to obtain
$$\phi(X) = \sum_{i=1}^p E_iXF_i,$$ where the $F_i$'s are the corresponding linear combinations of the $C_j$'s.

Remarkably, at this stage it is a theorem that the sets
$\{E_1,...,E_p \}$ and $\{F_1,...,F_p \}$ are linearly
independent, and hence this process terminates.

\bigskip

{\bf Step 4.} Given an invertible $S=(s_{i,j}) \in M_p$ with
inverse $S^{-1}=(t_{i,j}) \in M_p$, let $H_i = \sum_j s_{i,j}F_j,$
and $G_j = \sum_i t_{i,j}E_i.$ Then $$\|\phi\|_{cb} = \inf \Big\{
\|\sum_i G_iG_i^* \|^{1/2} \|\sum_i H_i^*H_i \|^{1/2} \Big\},$$
where the infimum is taken over all invertible matrices $S.$ It is
also enough to consider positive, invertible matrices for $S$.

This algorithm reduces the computation of $\|\phi\|_{cb}$ to a
series of matrix computations and only the last step might involve
a difficult minimization.

To begin to justify the algorithm, we begin with the last step.
First we show that $\phi(X) = \sum_i G_iXH_i.$ This can be seen
formally, because
\begin{multline*} \sum_i G_iXH_i = (G_1,...,G_p)\begin{pmatrix} X & 0
    & \dots & 0\\ 0 & X & \dots & 0 \\ \vdots & \vdots & \ddots &
    \vdots \\ 0 & 0 & \dots & X \end{pmatrix} \begin{pmatrix} H_1\\
    \vdots \\ H_p \end{pmatrix}  \\ =
  (E_1,...,E_p)(t_{i,j}I_n) \begin{pmatrix} X & 0 & \dots & 0 \\ 0 & X
    & \dots & 0 \\ \vdots & \vdots & \ddots & \vdots \\ 0 & 0 & \dots
    & X \end{pmatrix}(s_{i,j}I_n)
  \begin{pmatrix} F_1\\ \vdots \\ F_p \end{pmatrix}
\\ = \sum_i E_iXF_i = \phi(X), \end{multline*}
since the two scalar matrices commute past the direct sum in the
middle. Note that these scalar matrices behave like
``environmental operators'', that is, they are operators that act
exclusively on the environment of an open quantum system.

Next we need to see that the linear maps from $M_n$ to $M_k$,
which we denote by $\cl L(M_n,M_k)$, can be identified with the
tensor product, $M_{k,n} \otimes M_{n,k}$ via the map that sends
an elementary tensor, $A \otimes B$ to the map $\phi(X) = AXB.$ It
is easily seen that this extends to a linear map, $\Gamma: M_{k,n}
\otimes M_{n,k} \to \cl L(M_n,M_k),$ that a simple dimension count
shows is one-to-one and onto (both spaces have dimension
$n^2k^2$).

We now endow $M_{k,n} \otimes M_{n,k}$ with a norm so that
$\Gamma$ will be an isometry when $\cl L(M_n,M_k)$ is endowed with
the cb-norm. By the CB representation theorem, we see that if we
define for $U \in M_{k,n} \otimes M_{n,k},$
$$\|U\|_h = \inf \Big\{ \| \sum_i A_iA_i^\dagger \|^{1/2} \|\sum_i B_i^\dagger B_i
\|^{1/2} \Big\},
$$ where the infimum is taken over all ways to represent $U= \sum_i
A_i \otimes B_i$ as a sum of elementary tensors, then we will have
that $\|U\|_h = \|\Gamma(U) \|_{cb}.$

The above tensor norm is called the {\it Haagerup tensor norm} in
honor of U. Haagerup who was the first to notice the above
identification.  We write $M_{k,n} \otimes_h M_{n,k}$ to denote
the tensor product endowed with this norm and note that we have
just proved that:

\begin{thm}[Haagerup] The map $\Gamma: M_{k,n} \otimes_h M_{n,k} \to
  CB(M_n,M_k)$ defined by $\Gamma(A \otimes B)(X) = AXB$ is an
  isometric isomorphism.
\end{thm}

Here we use $CB(M_n, M_k)$ to denote the space of linear maps from
$M_n$ to $M_k$ endowed with the completely bounded norm. The above
isomorphism was greatly extended in work of Haagerup and
Effros-Kishimoto to other identifications between spaces of
completely bounded maps and Haagerup tensor products.

The above theorem reduces the justification of the above algorithm
to showing that if $\phi= \Gamma(U),$ then the algorithm correctly
computes, $\|U\|_h$. The fact that this algorithm correctly
computes $\|U\|_h$ for any operator spaces is proven in \cite{BP}.
We outline the key ideas below. For this, we will need a few facts
about tensor products of vector spaces.

Recall that if $\cl V$ and $\cl W$ are vector spaces, then every
element of $\cl V \otimes \cl W$ is a finite sum of elementary
tensors. The least number of elementary tensors that can be used
to represent an element $u \in \cl V \otimes \cl W$ is called the
{\it rank of u} and is denoted by ${\rm rank}\,(u)$.

\begin{prop}\cite{BP} Let $u \in \cl V \otimes \cl W.$ If $u =
  \sum_{i=1}^p v_i \otimes w_i$ then $p = rank(u)$ if and only if $\{v_1,...,v_p \}$ is a linearly independent set and $\{w_1,...,w_p \}$ is a linearly independent set. Moreover, if $u = \sum_{i=1}^p x_i \otimes y_i$ is another way to represent $u$ as a sum of elementary tensors and $p=rank(u),$ then
$$span \{ v_1,...,v_p \} = span \{ x_1,...,x_p \}$$
and $$span \{w_1,...,w_p \} = span \{y_1,..., y_p \}.$$
\end{prop}

\begin{prop}\cite{BP} Let $u \in \cl V \otimes \cl W.$ If we apply Step 1 and
  Step 2 of the above algorithm to $u= \sum_{i=1}^m a_i \otimes b_i,$
  to obtain $u= \sum_{i=1}^p e_i \otimes f_i,$ then $\{e_1,...,e_p \}$
  and $\{ f_1,..., f_p \}$ will be linearly independent sets and hence
  $rank(u) =p.$
\end{prop}

These facts are easily proved by applying maps of the form $f
\otimes id_W$ and $id_V \otimes g,$ where $f$ and $g$ are linear
functionals to $u$.

The remainder of the proof of the justification of the algorithm
is to show that at each stage, removing the linear dependencies
among the elements in the sum for $u$ reduces the Haagerup norm.
This is best seen at each stage of the algorithm. Say at Step 1,
when we choose the basis, $\{C_1,...,C_l \}$ and express, $B_i =
\sum_j d_{i,j}C_j$, if we first polar decompose the matrix
$(d_{i,j}) = (w_{i,j})(p_{i,j})$ where $W=(w_{i,j})$ is an $m
\times l$ partial isometry and $P=(p_{i,j})$ is an invertible $l
\times l$ positive matrix, then we have that $B_i= \sum_i
w_{i,j}\tilde{C_j},$ with $\tilde{C_i} = \sum_j p_{i,j}C_j.$  In
this case, the set $\{ \tilde{C_1},..., \tilde{C_l} \}$ is another
basis for the span of $\{ C_1,...,C_l \}$ and $\sum_i
\tilde{C_i}^\dagger \tilde{C_i} = \sum_i C_i^\dagger C_i.$
Moreover, using this basis, we would obtain another representation
for $\phi(X) = \sum_{j=1}^l \tilde{D_j}X\tilde{C_j},$ where
$\tilde{D_j} = \sum_i w_{i,j}A_i.$ Again, since $P$ is invertible,
the span of $\{ \tilde{D_1},..., \tilde{D_l} \}$ is the same as
the span of $\{ D_1,...D_l \}.$ Moreover, since $W$ is a partial
isometry, one finds that $\sum_i \tilde{D_i}\tilde{D_i}^\dagger
\le \sum_i A_iA_i^\dagger.$ Thus, the infimum of the Haagerup norm
expression over all linear combinations of the $D_i$'s and
$C_i$,'s which is the same as the infimum over all linear
combinations of the $\tilde{D_i}$'s and $\tilde{C_i}$'s is
smaller.

This proves that the quantity defining the Haagerup tensor norm
(which is the same as the CB norm) must be attained when the
coefficients of the generalized Choi-Kraus representation are
linearly independent, and hence represented by some choice of
basis for $span \{E_1,...,E_p \}$ and $span \{F_1,...,F_p \}.$

\subsection{Example}

In quantum information, maps given by the difference of two
(distinct) unitary maps form the most elementary class of linear,
non-completely positive maps of interest. We next show how the
algorithm can be used to derive a simple geometric technique that
computes the exact stabilized norm for maps in this class. We note
this result can be derived from a technical result of Herrero
(\cite{Herr}, Theorem~3.31), which is proved using operator
theoretic machinery. Moreover, the result is also stated more
recently in \cite{AKN97} without proof. Our proof is new and
elementary and gives a good illustration of the algorithm at work.
By the unitary invariance of the cb/$\diamond$ norm, observe that
we can compute the norm of any map $\mathcal U - \mathcal V$ once
we know how to compute it for any map of the form $\mathcal U -
{\rm id}$.

\begin{thm}\label{thm:cbfamily01}
    Let $U \in M_n$ be a unitary operator on a finite dimensional Hilbert space
and let $\Phi: M_n \to M_n$ be given by $\Phi(X) = U X U^\dagger -
X$. Then $\left\| \Phi \right\|_{cb} = \left\| \Phi^\dagger
\right\|_{cb} = \left\| \Phi \right\|_\diamond = \left\|
\Phi^\dagger \right\|_\diamond$ is equal to the diameter of the
smallest closed disc that contains all of the eigenvalues of $U$.
\end{thm}

\begin{proof}
    If $U$ is a scalar multiple of $I$ then $\Phi \equiv 0$ so the result immediately follows. Thus, we will
    assume from here on that $U$ is not a scalar multiple of $I$. It then follows from Step~4 of the algorithm that
$$ \|\Phi\|_{cb} = \inf \left\{ \left\| (U^\dagger, I) \begin{pmatrix} a & b\\c & d
\end{pmatrix}^{-1}\right\| \left\| \begin{pmatrix} a & b\\c & d \end{pmatrix}
\begin{pmatrix} U \\ -I \end{pmatrix} \right\| \right\}$$
where the infimum is over all invertible $2 \times 2$ scalar
matrices.

Now let $\mathbf{v} = \begin{pmatrix} a & c \end{pmatrix}^{T}$ and
$\mathbf{w} = \begin{pmatrix} b & d \end{pmatrix}^{T}$ so that
\begin{align*}
\left\| \begin{pmatrix} a & b\\c & d \end{pmatrix}\begin{pmatrix}
U \\ -I \end{pmatrix} \right\|^2 &= \left\| \begin{pmatrix} aU-bI \\
cU - dI \end{pmatrix} \right\|^2 \\ &= \left\| (|a|^2+ |b|^2 +
|c|^2+ |d|^2)I -(\bar{a}b+ \bar{c}d)U^\dagger - (a\bar{b} +
c\bar{d})U \right\| \\ &= \left\| (\|\mathbf{v}\|^2 +
\|\mathbf{w}\|^2)I - 2Re(\langle \mathbf{v},\mathbf{w}\rangle U)
\right\|
\end{align*}
If we let $D = ad - bc$ be the determinant of the matrix, then a
similar calculation shows that
$$\left\|(U^\dagger,I)\begin{pmatrix}d & -b\\ -c & a \end{pmatrix}D^{-1} \right\|^2 = |D|^{-2} \left\|(\|\mathbf{v}\|^2 + \|\mathbf{w}\|^2)I - 2Re( \langle \mathbf{v},\mathbf{w}\rangle U) \right\|$$
Thus it follows that
$$ \|\Phi\|_{cb} = \inf \{|D|^{-1} \|(\|\mathbf{v}\|^2 + \|\mathbf{w}\|^2)I - 2Re( \langle \mathbf{v},\mathbf{w}\rangle U) \| \}$$
where the infimum is taken over all $2 \times 1$ complex vectors
$\mathbf{v}$ and $\mathbf{w}$.

Now it is clear that this minimum will be attained when
$\mathbf{v}$ and $\mathbf{w}$ are rotated such that $\min_i
\left\{ Re( \langle \mathbf{v},\mathbf{w}\rangle \lambda_{i})
\right\}$ is as large as possible (while keeping $\mathbf{v}$ and
$\mathbf{w}$ of fixed length), where $\lambda_{i}$ ranges over all
eigenvalues of $U$. Thus, since multiplying $\mathbf{w}$ by
$e^{i\alpha}$ will not change $|D|$, it follows that
$$ \|\Phi\|_{cb} = \inf \{|D|^{-1} \|(\|\mathbf{v}\|^2 + \|\mathbf{w}\|^2)I - 2Re( \langle \mathbf{v},\mathbf{w}\rangle e^{i\alpha}U) \| \}$$
where $\alpha$ is such that the minimum real part of the
eigenvalues of $e^{i\alpha}U$ is as large as possible. Define $r$
to be this largest minimum real eigenvalue part.

Now, similar to before, we can multiply $\mathbf{w}$ by
$e^{i\beta}$ so that $\left| \langle \mathbf{v},\mathbf{w}\rangle
\right| = \langle \mathbf{v},e^{i\beta}\mathbf{w}\rangle$, and so
it follows that
$$ \|\Phi\|_{cb} = \inf \{|D|^{-1} \|(\|\mathbf{v}\|^2 + \|\mathbf{w}\|^2)I - 2\left| \langle \mathbf{v},\mathbf{w}\rangle \right|rI \| \}$$
where the infimum is now taken over all $2 \times 1$ real vectors
$\mathbf{v}$ and $\mathbf{w}$. It is now clear that we can assume
without loss of generality that $\left\| \mathbf{v} \right\|^2 +
\left\| \mathbf{w} \right\|^2 = 1$. It also follows from some
simple algebra that, given any two vectors $\mathbf{v}$ and
$\mathbf{w}$ such that $\left\| \mathbf{v} \right\| \neq \left\|
\mathbf{w} \right\|$, the value within this infimum will be made
smaller by scaling $\mathbf{v}$ and $\mathbf{w}$ so that $\left\|
\mathbf{v} \right\|^2 = \left\| \mathbf{w} \right\|^2 =
\frac{1}{2}$.

It then immediately follows from expanding out the terms within
the infimum that this is equivalent to the following minimization
problem
\begin{align*}
    \|\Phi\|_{cb} =& \min \Big\{\frac{1 - 2\left| ab + cd \right|r}{|ad - bc|}\Big\}    \\
    \text{such that  } & a^2 + c^2 = b^2 + d^2 = \frac{1}{2}
\end{align*}
Now, if $r \leq 0$ then it is easy to see that this minimum is
equal to $2$ by setting $a = d = \frac{1}{\sqrt{2}}$ and $b = c =
0$. Thus, it only remains to prove the conjecture in the case when
$r > 0$. If $r > 0$ then it is clear that this minimization
problem is equivalent to the one we get if we remove the absolute
value bars in the numerator.

We now form the Lagrangian of this problem:
$$\Lambda = \frac{1 - 2\left( ab + cd \right)r}{|ad - bc|} + \lambda_1 \left( a^2 + c^2 - \frac{1}{2} \right) + \lambda_2 \left( b^2 + d^2 - \frac{1}{2} \right)$$
If we now set $\frac{\partial \Lambda}{\partial b} =
\frac{\partial \Lambda}{\partial d}$, we arrive at the equation
$$\langle \mathbf{v}|\mathbf{w} \rangle = ab + cd = 2\left( a^2 + c^2 \right)\left( b^2 + d^2 \right)r = \frac{r}{2}$$
This, however, implies that $\theta = \arccos (r)$, where $\theta$
is the angle between $\mathbf{v}$ and $\mathbf{w}$. Thus, this
problem is minimized by vectors $\mathbf{v}$ and $\mathbf{w}$ that
are each of length $\frac{1}{\sqrt{2}}$ and separated by an angle
$\arccos (r)$. This, however, implies that $\left| D \right| =
\left\| \mathbf{v} \right\| \left\| \mathbf{w} \right\|
\sin{\theta} = \frac{1}{2}\sqrt{1 - r^2}$. Plugging this and $ab +
cd = \frac{r}{2}$ into the formula to be minimized, we learn that
$$\|\Phi\|_{cb} = 2\sqrt{1 - r^2}$$.

It now is a simple geometric argument that finally shows that this
value is equal to the diameter of the smallest closed disc
enclosing the eigenvalues of $U$, completing the proof.
\end{proof}

\section{Implementation of the Algorithm}

The implementation of the algorithm {\it via} Maple that we now
present is split up into several procedures, which will be
described as we present their code. We have kept variable names
within the code as close as possible to their counterparts
presented in the above theoretical discussion of the algorithm.
After the code has been presented and briefly explained, we
discuss its efficiency and provide an example that shows how the
code is used, comparing the results of the algorithm to known
theoretical results.

The first procedure, \verb|RandomPositive|, generates a random
positive matrix with eigenvalues in the interval
$(\verb|evalLower|,\verb|evalUpper|]$. This is achieved by
generating a diagonal matrix with entries contained in that
interval and conjugating by a random unitary. The random unitary
matrix is constructed by generating a matrix with random entries
from the square with corners at $0$ and $1 + i$ and then using the
Gram-Schmidt process on its columns.

\medskip
\begin{verbatim}
> RandomPositive := proc(ndim,evalLower,evalUpper)
    local r,i,RD,RM,V,W,U,P:r:=0:
    RD:=RandomMatrix(ndim,ndim,generator=evalLower+
      DBL_EPSILON..evalUpper,outputoptions=[shape=diagonal]):

    while r < ndim do
      RM:=RandomMatrix(ndim,ndim,generator=0.0..1.0) +
        I*RandomMatrix(ndim,ndim,generator=0.0..1.0):
      r:=round(Rank(RM)):
    od:
    for i from 1 to ndim do
      V[i]:=Vector(ndim,(j) -> RM[j,i]):
    od:
    W:=GramSchmidt(convert(V,list),normalized):
    U:=Matrix(ndim,ndim,(i,j) -> W[j][i]):
    P[1]:=MatrixMatrixMultiply(MatrixMatrixMultiply(U,RD),
      HermitianTranspose(U)):
    P[2]:=MatrixMatrixMultiply(MatrixMatrixMultiply(U,
      MatrixInverse(RD)),HermitianTranspose(U)):
    RETURN(P[1],P[2]):
  end:
\end{verbatim}

\medskip
The procedure \verb|IsCPMap| determines whether or not the
completely bounded map that it is given is actually a completely
positive map by determining whether or not its Choi matrix is
positive. This procedure is optional for the algorithm; it simply
serves to allow the algorithm to compute the completely bounded
norm of completely positive maps more quickly and more accurately
than it otherwise could.

\medskip
\begin{verbatim}
> IsCPMap := proc(CelA,CelB,NumOps,n,k)
    local i,j,x,Choi,LgChoi,MtxUnit:

    if not (n = k) then
      RETURN(false):
    else
      for i from 1 to n do
        for j from 1 to n do
          MtxUnit[i][j]:=OuterProductMatrix(UnitVector(i,n),
            UnitVector(j,n)):
          Choi[i][j]:=add(MatrixMatrixMultiply(
            MatrixMatrixMultiply(A[x],MtxUnit[i][j]),B[x]),
            x=1..NumOps):
        od:
      od:

      LgChoi:=Matrix(n^2,n^2,(i,j) -> Choi[floor((i-1)/n)+1]
        [floor((j-1)/n)+1][((i-1) mod n)+1,((j-1) mod n)+1]):
      RETURN(IsDefinite(LgChoi)):
    fi:
  end:
\end{verbatim}

\medskip
The procedures \verb|MakeLinIndep| and \verb|CellMatricize|
jointly perform steps 1 through 3 of the algorithm. They are
reasonably straightforward.

\medskip
\begin{verbatim}
> MakeLinIndep := proc(CelA,CelB,NumOps,n,k)
    local BM,BS,u,x,i,j,bg,BG,d,C,DOp,CM,CS,v,cg,CG,c,E,F:
    BM:=CellMatricize(CelB,NumOps,n,k):u:=round(Rank(BM)):
    BS:=Basis({Row(BM,[1..NumOps])}):
    for x from 1 to NumOps do
      for i from 1 to n*k do
        for j from 1 to u do
          bg[i,j] := BS[j][i]:
        od:
        bg[i,u+1] := BM[x,i]:
      od:
      BG[x]:=ReducedRowEchelonForm(
        Matrix(n*k,u+1,(i,j) -> bg[i,j])):

      for j from 1 to u do
        d[x,j]:=BG[x][j,u+1]:
      od:
    od:

    for x from 1 to u do
      C[x]:=Matrix(k,n,(i,j) ->
        add(d[l,x]*CelA[l][i,j],l=1..NumOps)):
      DOp[x]:=Matrix(n,k,(i,j) -> BS[x][j+(i-1)*k]):
    od:

    CM:=CellMatricize(C,u,k,n):v:=round(Rank(CM)):
    CS:=Basis({Row(CM,[1..u])}):
    for x from 1 to u do
      for i from 1 to n*k do
        for j from 1 to v do
          cg[i,j] := CS[j][i]:
        od:
        cg[i,v+1] := CM[x,i]:
      od:
      CG[x]:=ReducedRowEchelonForm(
        Matrix(n*k,v+1,(i,j) -> cg[i,j])):

      for j from 1 to v do
        c[x,j]:=CG[x][j,v+1]:
      od:
    od:

    for x from 1 to v do
      E[x]:=Matrix(k,n,(i,j)->CS[x][j+(i-1)*n]):
      F[x]:=Matrix(n,k,(i,j)->add(c[l,x]*DOp[l][i,j],l=1..u)):
    od:
    RETURN(E,F,v):
  end:

> CellMatricize := proc(Cel,a,b,c)
    RETURN(Matrix(a,b*c,(i,j) ->
      Cel[i][floor((j-1)/c)+1,j-c*floor((j-1)/c)])):
  end:
\end{verbatim}

\medskip
The procedure \verb|CBNorm| is the main procedure; it calls upon
the other procedures to compute the CB norm of the given map. The
procedure begins by determining whether or not the given map is
completely positive, and if so, returns the map's exact CB norm,
using the result of Theorem~\ref{thm:main}. If the map is not
completely positive, the algorithm described above begins to run.

Steps 1 through 3 are performed via the \verb|MakeLinIndep|
procedure. The minimization in step 4 is approximated by calling
upon \verb|RandomPositive| repeatedly to compute random positive
matrices with eigenvalues in the interval $(0,1]$ and taking the
minimum of the resulting norm estimates. We can restrict ourselves
to using only positive matrices with eigenvalues in the interval
$(0,1]$ rather than all positive matrices because multiplying such
a matrix by a constant will not change the resulting norm
estimate.

The inputs to the function are \verb|CelA|, an array of the map's
$A$ operators in one of its generalized Choi-Kraus
representations; \verb|CelB|, an array of the map's $B$ operators
in the same representation; \verb|NumIts|, the number of random
matrices to be used to estimate the norm (a higher number will
produce a more accurate estimate but will take longer to compute);
and \verb|NumOps|, the number of $A$ and $B$ operators that there
are for the map in its given representation.

\medskip
\begin{verbatim}
> CBNorm := proc(CelA,CelB,NumIts,NumOps)
    local i,x,n,k,CBGuess,NewCBGuess,ST,G,H,GG,HH,GGCB,HHCB,EF:
    n:=ColumnDimension(A[1]):k:=RowDimension(A[1]):
    CBGuess:=infinity:

    if IsCPMap(CelA,CelB,NumOps,n,k) then
      CBGuess:=Norm(add(MatrixMatrixMultiply(
        MatrixMatrixMultiply(A[x],IdentityMatrix(k)),B[x]),
        x=1..NumOps),2):
    else
      EF:=MakeLinIndep(CelA,CelB,NumOps,n,k):

      for i from 1 to NumIts do
        ST:=RandomPositive(EF[3],0,1):

        for x from 1 to EF[3] do
          H[x]:=Matrix(n,k,(i,j) ->
            add(ST[1][x,l]*EF[2][l][i,j],l=1..EF[3])):
          G[x]:=Matrix(k,n,(i,j) ->
            add(ST[2][l,x]*EF[1][l][i,j],l=1..EF[3])):
          HH[x]:=MatrixMatrixMultiply(
            HermitianTranspose(H[x]),H[x]):
          GG[x]:=MatrixMatrixMultiply(
            G[x],HermitianTranspose(G[x])):
        od:

        HHCB:=simplify(Matrix(k,k,(i,j) ->
          add(HH[l][i,j],l=1..EF[3]))):
        GGCB:=simplify(Matrix(k,k,(i,j) ->
          add(GG[l][i,j],l=1..EF[3]))):
        NewCBGuess:=Re(sqrt(evalf(Norm(GGCB,2)) *
          evalf(Norm(HHCB,2)))):
        if NewCBGuess < CBGuess then
          CBGuess:=NewCBGuess:
        fi:
      od:
    fi:
    RETURN(CBGuess):
  end:
\end{verbatim}

\medskip
To use the provided code, run the command
\verb|with(LinearAlgebra):| in Maple and then load in the
procedures defined above. As an illustration we return to a
special case of the class of maps discussed above.

\begin{exam}
Let $U$ be the $3\times 3$ diagonal unitary matrix with
eigenvalues
$e^{\left(\frac{5i\pi}{4}\right)},e^{\left(i\pi\right)},\text{ and
}e^{\left(\frac{3i\pi}{4}\right)}$. Then the following code
computes the CB norm of the map $\phi\left( X \right) =
U^{\dagger}XU - X$.

\medskip
\begin{verbatim}
> NumOps:=2:
> NumIts:=100:
> A:=Array(1..NumOps):
> B:=Array(1..NumOps):

> A[1]:=DiagonalMatrix([exp(-5*I*Pi/4),exp(-I*Pi),
exp(-3*I*Pi/4)]):
> A[2]:=IdentityMatrix(3):
> B[1]:=DiagonalMatrix([exp(5*I*Pi/4),exp(I*Pi),
exp(3*I*Pi/4)]):
> B[2]:=-IdentityMatrix(3):

> CBNorm(A,B,NumIts,NumOps);
\end{verbatim}

\medskip
Running this code gives an output of $1.449$ in just under $7$
seconds. Theorem~\ref{thm:cbfamily01} tells us however that
$\left\| \phi \right\|_{cb} = \sqrt{2}$, so our algorithm is
correct to two significant digits. To get a more accurate
estimate, we can of course increase the number of iterations from
$100$, and it should be clear how to modify this code to find the
CB norm of other maps.
\end{exam}

\subsection{Efficiency}

To look at the efficiency of the algorithm, we assume that $n = k$
and consider Steps 1 - 3 separately from Step 4, as Steps 1 - 3
need only be run once for a given map, while we may wish to run
our implementation of Step 4 hundreds or thousands of times for
the same map.

First note that the most efficient algorithm that could possibly
exist for computing the CB norm of a general CB map is $O(n^4)$,
which can be seen by observing that a general CB map will have
about $n^2$ linearly independent generalized Choi-Kraus operators,
each of which has $n^2$ entries that must each be read at least
once. One can observe, however, that the efficiency of Steps 1 - 3
of this algorithm is $O(n^8)$, as we need to apply Gaussian
Elimination to about $n^2$ matrices each of dimension about $n^2
\times n^2$.

If we know already that the generalized Choi-Kraus operators of
our given representation are linearly independent, we can simply
proceed to Step 4 of the algorithm, which has efficiency that can
be seen to be $O(n^5)$, as the step of computing $\sum_i G_iG_i^*$
and $\sum_i H_i^*H_i$ from the families of matrices $\left\{ G_i
\right\}$ and $\left\{ H_i \right\}$ is $O(n^5)$ (if computed
using the standard matrix multiplication algorithm). Since all of
the other operations in Step 4 are at least as efficient as
$O(n^4)$, it follows that if we can find a clever way to compute
$\sum_i G_iG_i^*$ and $\sum_i H_i^*H_i$, we can reduce the overall
order of this step.

One of the most obvious ways to improve the order of Step 4 is to
use an algorithm like the Strassen algorithm or the
Coppersmith-Winograd algorithm \cite{CW} to perform our matrix
multiplications. Employing these matrix multiplication techniques
would then reduce the order of Step 4 to about $O(n^{4.807})$ or
$O(n^{4.376})$, respectively. However, the resolution of either of
two conjectures in \cite{CUKS} would imply that general matrix
multiplication can be carried out in about $O(n^2)$ time, which
would imply that Step 4 can be carried out in about $O(n^4)$ time,
making our algorithm optimal for maps in which we already have a
linearly independent representation.

\strut

{\noindent}{\it Acknowledgements.} The genesis for this paper was
the talk of the third named author, and subsequent positive
participant reaction, at the February 2007 Banff Workshop on
Operator Structures in Quantum Information Theory. We gratefully
acknowledge BIRS for its hospitality, and the workshop
participants for making it a very fruitful meeting. We are also
grateful to Renato Renner, Marcus Silva, Robert Spekkens, and John
Watrous for helpful conversations. N.J. was partially supported by
NSERC CGS, V.I.P. was partially supported by NSF grant
DMS-0600191, and D.W.K. was partially supported by NSERC, ERA,
CFI, and ORF.

\end{document}